\documentclass[final,5p,times,twocolumn]{elsarticle}
\usepackage[T1]{fontenc}
\usepackage[utf8]{inputenc}

\usepackage{amsmath}
\usepackage{amsfonts}
\usepackage{amssymb}
\usepackage{amsxtra}
\usepackage{array}
\usepackage{color}
\usepackage{dcolumn}
\usepackage{graphicx}
\usepackage{hepunits}
\usepackage{xspace}
\usepackage{CJKutf8}

\usepackage{subcaption}
\RequirePackage{comment}

\definecolor{purple}{rgb}{0.5,0,0.5}
\definecolor{blue}{rgb}{0.0,0,0.9}
\definecolor{prdblue}{rgb}{0.133,0.118,0.498}
\usepackage[colorlinks=true, pdfstartview=FitV, linkcolor=prdblue, citecolor= prdblue, urlcolor=prdblue]{hyperref}

\usepackage[mathscr,scaled=1.15]{urwchancal}
\DeclareFontFamily{OT1}{pzc}{}
\DeclareFontShape{OT1}{pzc}{m}{it}%
{<-> s * [1.15] pzcmi7t}{}
\DeclareMathAlphabet{\mathpzc}{OT1}{pzc}{m}{it}

\biboptions{sort&compress}

\journal{Physics Letters B}

\begin{document}
\begin{CJK*}{UTF8}{gbsn}

\begin{frontmatter}

\title{$\,$\\[-7ex]\hspace*{\fill}{\normalsize{\sf\emph{Preprint no}. NJU-INP 126/26}}\\[1ex]
Perspective on Proton Transversity Distribution Functions}

\author[NJU,INP]{Ke-Xin Zeng (曾可欣)%
        $\,^{\href{https://orcid.org/0009-0001-6575-5645}{\textcolor[rgb]{0.00,1.00,0.00}{\sf ID}},}$}
       
\author[WuHu]{Peng Cheng (程鹏),%
    $^{\href{https://orcid.org/0000-0002-6410-9465}{\textcolor[rgb]{0.00,1.00,0.00}{\sf ID}}\,,}$}

\author[ECJU]{Yang Yu (俞杨)%
       $\,^{\href{https://orcid.org/0009-0008-8011-3430}{\textcolor[rgb]{0.00,1.00,0.00}{\sf ID}}\,,}$}
       
\author[ECT]{Daniele Binosi%
    $^{\href{https://orcid.org/0000-0003-1742-4689}{\textcolor[rgb]{0.00,1.00,0.00}{\sf ID}}\,,}$}

\author[NJU,INP]{Craig D.\ Roberts%
       $^{\href{https://orcid.org/0000-0002-2937-1361}{\textcolor[rgb]{0.00,1.00,0.00}{\sf ID}},}$}

\address[NJU]{
School of Physics, \href{https://ror.org/01rxvg760}{Nanjing University}, Nanjing, Jiangsu 210093, China}
\address[INP]{
Institute for Nonperturbative Physics, \href{https://ror.org/01rxvg760}{Nanjing University}, Nanjing, Jiangsu 210093, China}

\address[WuHu]{
Department of Physics, \href{https://ror.org/05fsfvw79}{Anhui Normal University}, Wuhu, Anhui 24100, China}

\address[ECJU]{School of Science, \href{https://ror.org/05x2f1m38}{East China Jiaotong University}, Nanchang, Jiangxi 330013, China}

\address[ECT]{European Centre for Theoretical Studies in Nuclear Physics
            and Related Areas  (\href{https://ror.org/01gzye136}{ECT*})\\ Villa Tambosi, Strada delle Tabarelle 286, I-38123 Villazzano (TN), Italy
\\[1ex]
\href{mailto:cdroberts@nju.edu.cn}{cdroberts@nju.edu.cn} (CDR)
\\[1ex]
Date: 2026 November 17\\[-6ex]. 
}

\begin{abstract}
Using available information on proton tensor charges and constraints from perturbative quantum chromodynamics (QCD), \emph{Ans\"atze} are introduced for hadron-scale in-proton valence-quark transversity distributions functions (DFs).  
The concept of a QCD effective charge, which defines an evolution scheme for collinear parton distribution and fragmentation functions that is all-orders exact, is then used to deliver predictions for the $x$-dependence of these DFs at the scale $\zeta_{2} = 2\,$GeV.
Although the approach is largely algebraic, its predictions remain viable after comparisons with results obtained using computationally intensive methods.  
Thus, given the scarcity of data that can today be used for inference of proton transversity DFs, it may be worth using the results herein to inform and constrain related experiments and phenomenological fits.  
\end{abstract}


\end{frontmatter}
\end{CJK*}

\section{Introduction}
The future of strong interaction physics is defined by a programme bent on explaining the origin of hadron mass and spin \cite{NAP25171}.  
In this connection, Poincar\'e-invariant quantities, such as hadron elastic and transition form factors, are of special value because methods for their extraction from data are mature; see, \emph{e.g}., Refs.\,\cite{Aguilar:2019teb, Brodsky:2020vco, Achenbach:2025kfx, Wojtsekhowski:2026zxn}.  
Notwithstanding, in addressing hadron mass and spin, much of the discourse is conducted in terms of 
parton distribution functions (DFs).
They draw their appeal from connections with the parton model \cite{Feynman:1969ej}, which was a crucial step in the development of quantum chromodynamics (QCD) \cite{Fritzsch:1973pi}.

\begin{figure}[t]
\begin{center}
\begin{tabular}{l}
\vspace*{2ex}\\
\hspace*{0.\textwidth}{\sf A - helicity} \\[-6ex]
\hspace*{14ex}\includegraphics[clip, width=0.27\textwidth]{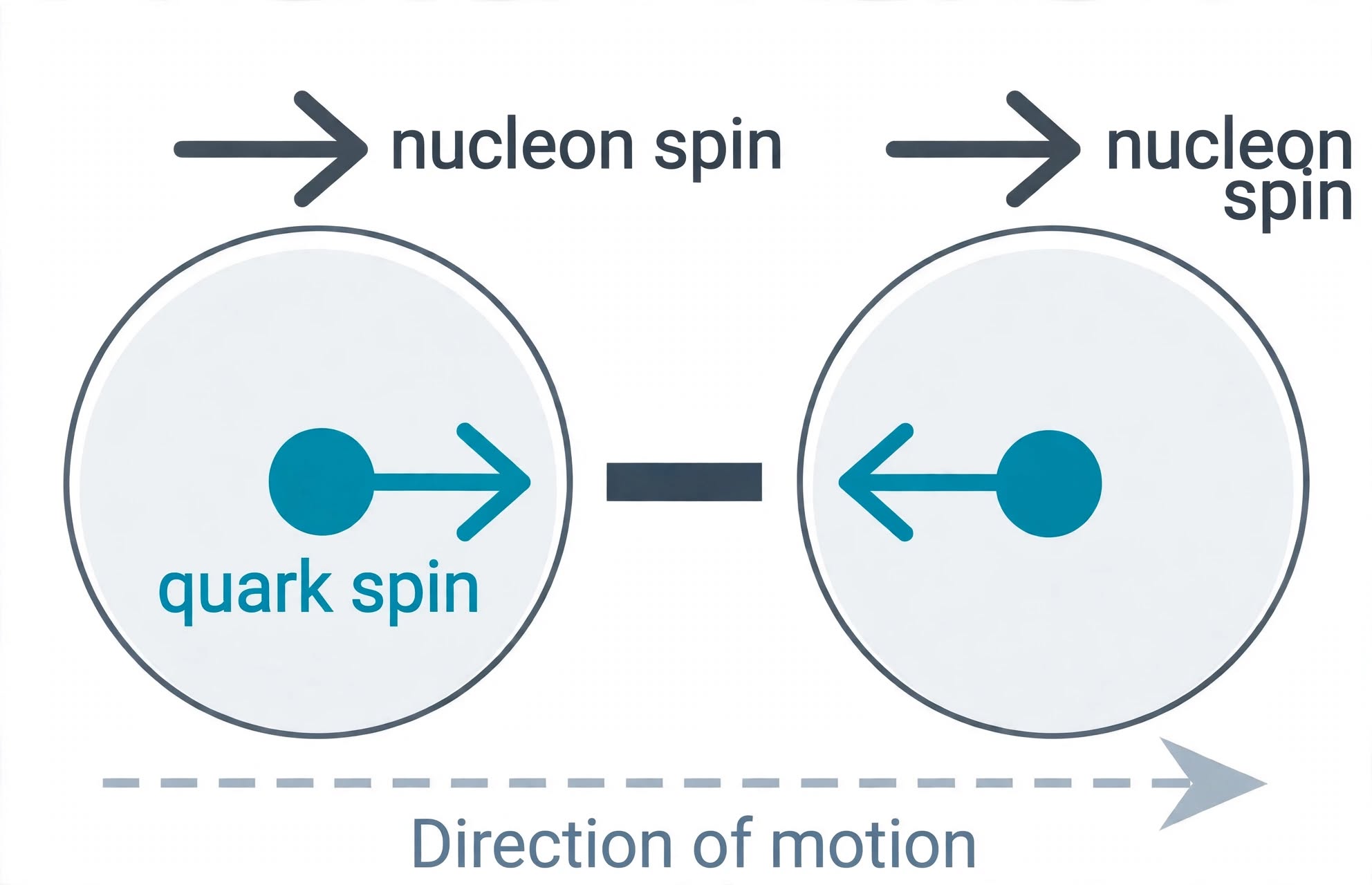} \\
\vspace*{2ex}\\
\hspace*{0.\textwidth}{\sf B - transversity} \\[-6ex]
\hspace*{14ex}\includegraphics[clip, width=0.27\textwidth]{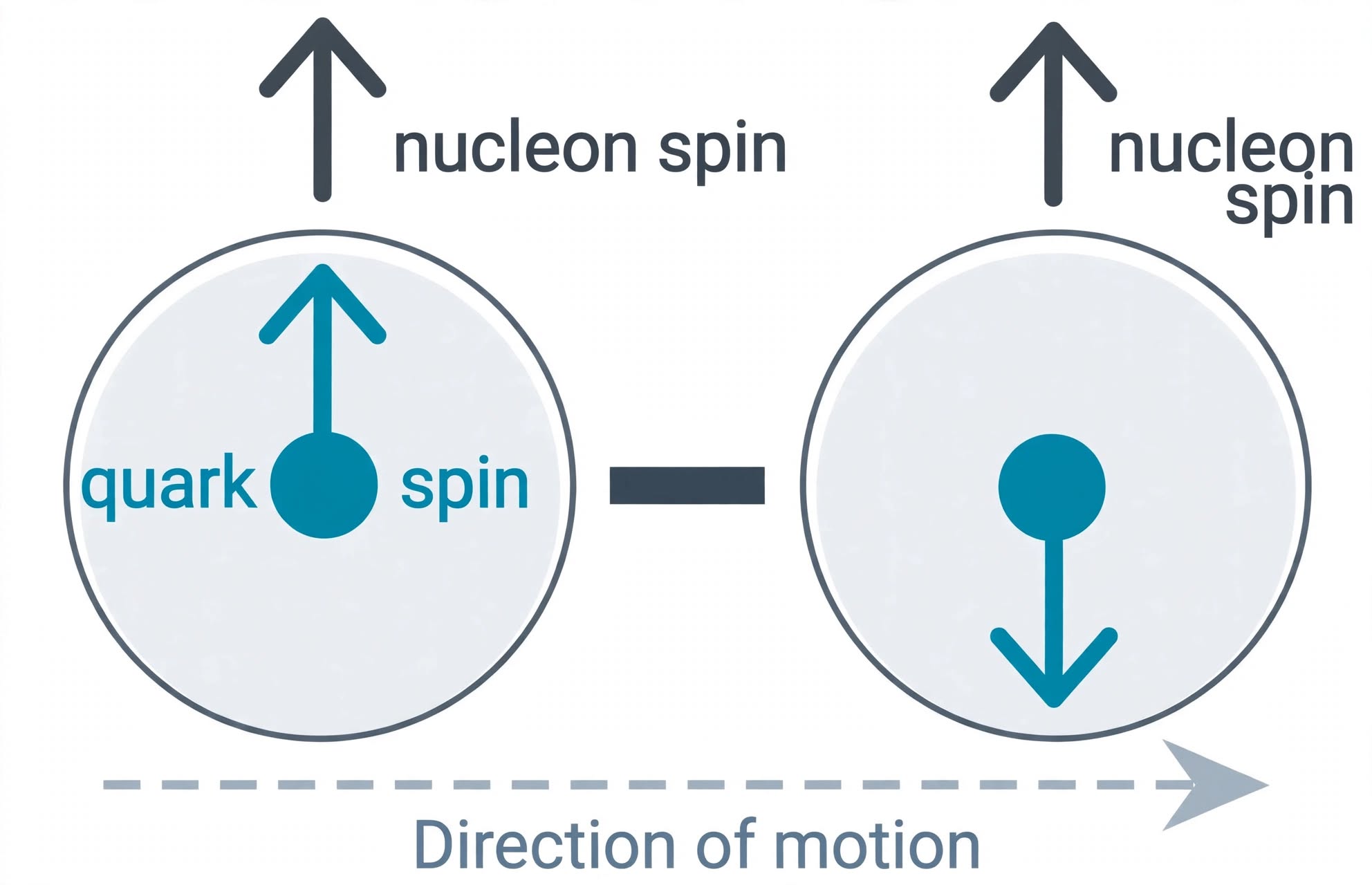} 
\end{tabular}
\end{center}
\vspace*{-4ex}

\caption{\label{FDFs}
A $J=1/2$ bound-state has two spin-dependent collinear valence-quark DFs: 
{\sf Panel A} -- helicity DF, $\Delta q(x)$,  measures the light-front number-density of quarks with longitudinal polarisation parallel to that of the bound-state minus that of quarks with antiparallel polarisation; 
{\sf Panel B} -- transversity DF, $h_1(x)$ -- here, longitudinal is replaced by transverse polarisation.
}
\end{figure}

In principle, collinear DFs, which are related to overlaps of light-front wave functions \cite{Diehl:2000xz}, describe the probability of finding a given in-hadron partonic degree of freedom (dof) carrying a certain fraction of a specific hadron property when that parton possesses a light-front fraction $x$ of the hadron's total momentum.  These DFs depend on the energy scale, $\zeta$, of the probe used to resolve the hadron.
Empirical results are typically extracted from data obtained in high-energy scattering experiments.  
High energy is necessary because one must use a kinematic domain whereupon the measurable cross section can reliably be approximated as a convolution.  
On such a domain, the convolution kernel is supposed to be target-independent and calculable in perturbative QCD.  
However, even today, discussions continue over whether or not the kernels in common use supply a sound approximation \cite{Cui:2021mom, Wang:2025usl}. 

Three types of collinear DF are associated with proton structure studies.
Each is connected with a light-front transverse momentum ($k_\perp$) dependent DF (TMD) \cite[Sec.\,2.9]{Boussarie:2023izj}.
The spin-independent DFs can be related to questions surrounding the origin of hadron mass; see, \emph{e.g}., Refs.\,\cite{Ji:1995sv, Lorce:2018egm, Cui:2021mom, Roberts:2022rxm}.

Two collinear DFs are connected with proton spin, \emph{viz}.\  helicity, $\Delta {\mathpzc p}(x;\zeta)$ and transversity, $h_1^{\mathpzc p}(x;\zeta) \equiv \delta_T {\mathpzc p}(x;\zeta)$, where $\mathpzc p$ represents one or another in-proton parton species.
Interpretations of such valence quark DFs are sketched in  Fig.\,\ref{FDFs}.

Helicity distributions have received much attention since experiments at CERN \cite[EMC]{EuropeanMuon:1987isl}, forty years ago; and interest in transversity DFs, too, began soon after those experiments.
Helicity DFs are associated with chiral even operators; hence, are readily accessible via deep inelastic scattering (DIS) experiments.  
They connect DIS with nucleon axial charges \cite{Bjorken:1969mm}.
Whilst these nucleon properties are conserved, their dynamically determined values are not constrained by symmetries. 

Transversity, on the other hand, is chiral-odd, making it more difficult to extract.
One needs to use processes in which $h_1$ is combined with some other aspect of the interaction that also involves a chiral-odd quantity.  This complication typically means that a collection of measurements must be analysed in combination, \emph{e.g.}, dihadron production in $e^+e^-$ collisions, semi-inclusive DIS from a transversely polarised target, and proton-proton collisions, with one proton being transversely polarised \cite{Radici:2018iag, Cocuzza:2023oam}.  
The integrated strengths of proton transversity DFs measure its tensor charges.  
They are not conserved; yet, are important dynamically determined nucleon qualities, which have widespread relevance, \emph{e.g}., in searches for physics beyond the Standard Model using the neutron electric dipole moment \cite{Bhattacharya:2015esa}.

It is worth noting that the valence dof helicity and transversity distributions are identical in nonrelativistic systems because, in such systems, boosts and rotations would commute with the strong interaction Hamiltonian.  In principle, this equivalence is broken in QCD, but it is natural to ask: just how great are the differences?  We address that issue herein using a unified, primarily algebraic approach within the framework of continuum Schwinger function methods (CSMs) \cite{Eichmann:2016yit, Burkert:2017djo, Ding:2022ows, Mezrag:2023nkp}.  
The analysis is not limited to that question, however.
In delivering results for both types of distributions, it can also be of value in assessing phenomenological inferences thereof, which today remain subject to significant systematic uncertainties; see, \emph{e.g}., Refs.\,\cite{Bertone:2024taw, Cruz-Martinez:2025ahf, Shahrzad:2026fgv, Radici:2018iag, Cocuzza:2023oam}.

\section{Effective Charge}
We proceed following the approach to helicity DFs employed in Ref.\,\cite{Cheng:2023kmt}.
The starting point is the theory of QCD effective charges, introduced in Refs.\,\cite{Grunberg:1980ja, Grunberg:1982fw} and reviewed in Ref.\,\cite{Deur:2023dzc}.
This framework enables one to identify an effective charge (running coupling), $\alpha_{1\ell}(k^2)$, which, when used to integrate the leading-order DGLAP equations \cite{Dokshitzer:1977sg, Gribov:1971zn, Lipatov:1974qm, Altarelli:1977zs}, defines an evolution scheme for \emph{all} collinear parton DFs and fragmentation functions (FFs) that is both all-orders exact and applicable at all scales \cite{Yin:2023dbw, Xing:2025eip}.
In this way, one arrives at a nonperturbative extension of DGLAP evolution.  

Details about the form of $\alpha_{1\ell}(k^2)$ are largely irrelevant; for (almost) all purposes, one need only know that such a charge exists.  
Nevertheless, the QCD process-independent (PI) running coupling, defined in Ref.\,\cite{Binosi:2016nme} and calculated in Ref.\,\cite{Cui:2019dwv}, has each required property.
With such a charge in hand, it follows that there is a hadron scale, $\zeta_{\cal H}$, whereat every property of a given hadron is carried by its valence degrees of freedom (dofs), \emph{viz}.\ momentum, spin, etc.
At $\zeta_{\cal H}$, DFs associated with glue and sea quarks are zero.
In standard implementations \cite{Yin:2023dbw}, it has proven successful to take the same value of $\zeta_{\cal H}$ for every hadron; see, \emph{e.g}., Refs.\,\cite{Lu:2022cjx, Xu:2024nzp, Zeng:2026peb, Xing:2026rnl}.

This effective charge scheme is called AO evolution.
The universality of the approach -- all collinear DFs and FFs, all scales, and all hadrons -- makes the AO scheme predictive. 

\begin{table}[t]
\caption{\label{anValence}
Gegenbauer coefficients that define the hadron-scale valence quark DFs in Eq.\,\eqref{ExpandGegenbauer}.
Each entry should be divided by $10^3$.
 }
%
\begin{center}
\begin{tabular*}
{\hsize}
{
c@{\extracolsep{0ptplus1fil}}
c@{\extracolsep{0ptplus1fil}}
c@{\extracolsep{0ptplus1fil}}
c@{\extracolsep{0ptplus1fil}}
c@{\extracolsep{0ptplus1fil}}}\hline\hline
\multicolumn{5}{c}{$a_n^{\mathpzc u}$}\\\hline
1 & 2 & 3 & 4 & 5 \\
$403$ & $112$ & $7.31 $ & $-10.4$ & $-4.90\ $ \\\hline
6 & 7 & 8 & 9 & 10 \\
$-0.0474$ & $1.15$ & $0.828$ & $0.334$  & $0.0635$ \\\hline\hline
\multicolumn{5}{c}{$a_n^{\mathpzc d}$}\\\hline
1 & 2 & 3 & 4 & 5 \\
$482$ & $161$ & $20.8$ & $-11.5$ & $-7.11\ $ \\\hline
6 & 7 & 8 & 9 & 10 \\
$-0.430$ & $1.50$ & $1.03$ & $0.349$  & $0.0543$ \\\hline\hline
\end{tabular*}
\end{center}
\end{table}

Having established the existence of $\zeta_{\cal H}$, one can discuss the properties of valence dof DFs at this scale.  
We take the unpolarised $u$ and $d$ valence quark DFs calculated in Ref.\,\cite{Chang:2022jri}.
They were obtained using a quark + fully-interacting diquark, $q(qq)$, picture of nucleon structure, in which the nucleon is described by a  Poincar\'e-covariant Faddeev wave function that is dominated by isoscalar-scalar (SC) diquark correlations (65\% of the canonical normalisation integral) yet contains material isovector-axialvector (AV) components (whose presence contributes the remaining 35\%); see, e.g., Ref.\,\cite{Liu:2022ndb}.
These unpolarised DFs are reliably interpolated by the following finite sum of Gegenbauer polynomials:
\begin{equation}
\label{ExpandGegenbauer}
{\mathpzc q}(x;\zeta_{\cal H}) = n_{\mathpzc q} 140 x^3(1-x)^3 \left[1 + \sum_{n=1}^{10} a_n^{\mathpzc q} C_n^{7/2}(1-2x)\right]\,,
\end{equation}
${\mathpzc q}={\mathpzc u}, \mathpzc d$, $n_{\mathpzc u}=2n_{\mathpzc d}=2$, where the coefficients are listed in Table~\ref{anValence}.
In addition to being phenomenologically efficacious, these DFs compare favourably with available results from lattice-regularised QCD (lQCD); see, \emph{e.g.}, Ref.\,\cite[Fig.\,1]{Yu:2026poj}. 

Using the same $q(qq)$ picture, one finds that the nucleon axial charge is $g_A=1.25(3)$ \cite{ChenChen:2022qpy}, where the uncertainty reflects that on the SC and AV diquark masses: $m_{\rm SC}\approx 0.8\,$GeV; $m_{\rm AV} \approx 0.9\,$GeV.  
Furthermore, the $u$ quark fraction of the nucleon axial charge is $g_A^u/g_A=0.76(1)$; the $d$ quark ratio is $g_A^d/g_A=-0.24(1)$; and $g_A^d/g_A^u=-0.32(2)$. \label{gAvalues}
Notably, the singlet axial charge of the proton is \cite[Sec.\,9]{Ding:2022ows}: $a_0 = 0.65(2)$.
At $\zeta_{\cal H}$ in a Poincar\'e-covariant treatment of the problem, the remainder of the proton spin is invested in quark orbital angular momentum; see, \emph{e.g}., Ref.\,\cite{Cheng:2025sdp}.

\section{Helicity Distributions}
\label{SecHelicity}
To extend the analyses of unpolarised DFs in Ref.\,\cite{Chang:2022jri} to the helicity case, Ref.\,\cite{Cheng:2023kmt} followed Refs.\,\cite{Liu:2019vsn, Han:2021dkc} in exploiting constraints suggested by analyses in perturbative QCD \cite{Brodsky:1994kg} to develop \emph{Ans\"atze} for the hadron-scale valence-quark helicity DFs.
The foundation is the observation that one may, with complete generality, write
\begin{equation}
\Delta {\mathpzc q}(x;\zeta_{\cal H}) = {\mathpzc s}^{\mathpzc q}{\mathpzc r}^{\mathpzc q}(x) {\mathpzc q}(x;\zeta_{\cal H}) ,
\end{equation}
where ${\mathpzc s}^{\mathpzc u} = 1 = - {\mathpzc s}^{\mathpzc d}$ and 
${\mathpzc r}^{\mathpzc q}(x) $ is a smooth function.  For the helicity distributions, the assumption is that ${\mathpzc r}^{\mathpzc q}(x)$ is a monotonically increasing function that vanishes at $x=0$ and is nonzero and finite at $x=1$.  

\begin{table}[t]
\caption{\label{gammavalues}
Referring to Eq.\,\eqref{ratiofunctions}, central values of the mapping parameters, $\gamma^{\mathpzc q}$, that reproduce the given quark's contribution to the proton axial charge.
 } 
%
\begin{center}
\begin{tabular*}
{\hsize}
{
l@{\extracolsep{0ptplus1fil}}
c@{\extracolsep{0ptplus1fil}}
c@{\extracolsep{0ptplus1fil}}
c@{\extracolsep{0ptplus1fil}}
c@{\extracolsep{0ptplus1fil}}}\hline\hline
$i$ & 1 & 2 & 3 & 4 \\\hline
$\gamma_i^{\mathpzc u}\ $ & $0.350$ & $0.621$ & $ 0.575$ & $0.853\ $\\
$\gamma_i^{\mathpzc d}\ $ & $1.47\phantom{0} $ & $1.26\phantom{1} $ & $2.62\phantom{5} $ & $1.60\phantom{3}\ $ \\[+0.5ex]
\hline\hline
\end{tabular*}
\end{center}
\end{table}

\begin{figure}[t]
\centerline{\includegraphics[width=0.44\textwidth]{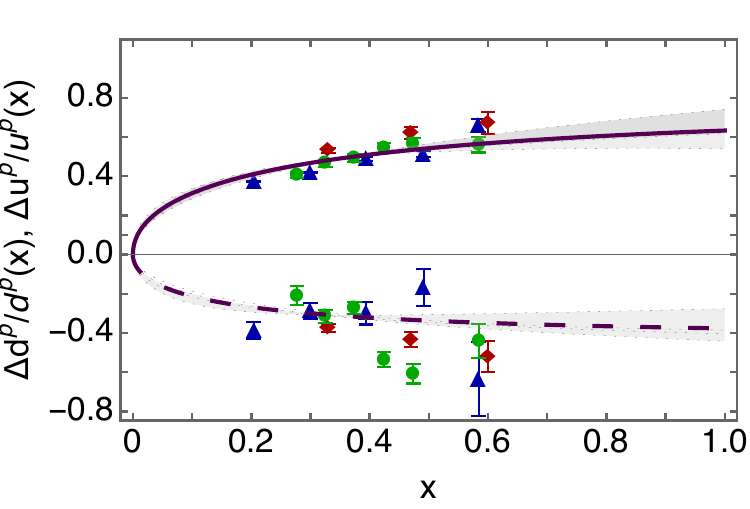}}
\caption{\label{FigDeltaud}
Legend.
Predicted ratios $\Delta {\mathpzc q}(x;\zeta_{\cal H}) /{\mathpzc q}(x;\zeta_{\cal H})$, obtained using Eq.\,\eqref{ratiofunctions} and Table~\ref{gammavalues}.
The uncertainty expresses the variation over choices for $\gamma^{\mathpzc q}$ and the ratios are scale invariant under AO evolution.  
Upper panel for $\mathpzc u$ and lower panel for $\mathpzc d$.
Data:  
red diamonds \cite[E99-117]{JeffersonLabHallA:2004tea};
blue up-triangles \cite[CLAS EG1]{CLAS:2006ozz};
green circles \cite[E06-014]{JeffersonLabHallA:2014mam}.
}
\end{figure}

Four distinct such mappings were employed \cite{Cheng:2023kmt}:
\begin{subequations}
\label{ratiofunctions}
\begin{align}
r_1(x,\gamma) & = \sqrt{x}/[1+\gamma \sqrt{x}] \,, &
r_2(x,\gamma) & = \sqrt{x}/[\gamma+ \sqrt{x}] \,, \\
r_3(x,\gamma) & = \sqrt{x}/[1+ \gamma x] \,, &
r_4(x,\gamma) & = \sqrt{x}/[\gamma+ x] \,,
\end{align}
\end{subequations}
with $\gamma_i^{\mathpzc q}$ being fixed by requiring
$\int_0^1 dx\,\Delta{\mathpzc q}(x;\zeta_{\cal H}) = g_A^{\mathpzc q}$ in each case. 
This leads to the values listed in Table~\ref{gammavalues}, which deliver the flavour-separated ratios drawn in Fig.\,\ref{FigDeltaud} and their linked uncertainties.  
Given that the $\gamma^{\mathpzc q}$ values were fixed independently of the data included for comparison in Fig.\,\ref{FigDeltaud}, the evident agreement supports the straightforward \emph{Ansatz} adopted in Ref.\,\cite{Cheng:2023kmt}.

Working with these nucleon helicity DFs, Ref.\,\cite{Cheng:2023kmt} subsequently arrived at both a 
viable explanation of modern proton spin measurements \cite[COMPASS]{COMPASS:2016jwv} 
and an angular momentum decomposition of the proton spin, at $\zeta_2 = 2\,$GeV \cite{Yu:2024ovn, Yu:2025fer}: quark spin $\approx 45$\%, quark orbital angular momentum $\approx 15$\%, and gluon $\approx 40$\%.
These values align with contemporary lQCD estimates \cite{Alexandrou:2026lrh}.
Having thus established the efficacy of the approach, we now employ it for transversity distributions.

\section{Transversity Distributions: Hadron Scale}
\label{SecT}
In order to adapt the Ref.\,\cite{Cheng:2023kmt} scheme to transversity distributions, one must have values for the $u$ and $d$ quark contributions to the proton tensor charge.
The difference with respect to axial charges is that the relevant tensor current is not conserved; so, quark (and hadron) tensor charges are scale dependent.  

This is somewhat like the running of the quark current mass: $m(\zeta) \to 0$ as $\zeta\to \infty$; although in that case, the Lagrangian product $m(\zeta)\bar q(\zeta) q(\zeta)$ is constant.  Reexpressing mass running in the context of the AO scheme, one has:
\begin{subequations}
\begin{align}
\frac{m(\zeta)}{m(\zeta^\prime)} & = {\mathpzc E}_{\mathpzc V}(\zeta,\zeta^\prime ; 4) , \label{Eq4a}\\
{\mathpzc E}_{\mathpzc V}(\zeta,\zeta^\prime ; a)
& = \exp\bigg[
\tfrac{1}{4\pi}\int_{\zeta^2}^{\zeta^{\prime 2}} ds \frac{\alpha_{1\ell}(s)}{s} 
\bigg]^a.
\end{align}
\end{subequations}
For any value of $\Lambda_{\rm UV}$, Eq.\,\eqref{Eq4a} can be rewritten as follows:
\begin{equation}
m(\zeta)  {\mathpzc E}_{\mathpzc V}(\Lambda_{\rm UV} ,\zeta ; 4) = 
m(\zeta^\prime)  {\mathpzc E}_{\mathpzc V} (\Lambda_{\rm UV}, \zeta^\prime ; 4) ,
\end{equation}
where both sides of the equation are independent of the scale.  
Now making use of asymptotic freedom, one obtains the renormalisation group invariant (RGI) quark current mass \cite{Politzer:1976tv}:
\begin{equation}
\hat m = \lim_{\Lambda_{\rm UV} \to\infty}  \ln [ \Lambda_{\rm UV}^2/\Lambda_{\rm QCD}^2 ]^{\gamma_m} \, 
m(\zeta)  {\mathpzc E}_{\mathpzc V}(\Lambda_{\rm UV} ,\zeta ; 4) ,
\end{equation}
where the anomalous mass dimension is $\gamma_m = 12/(33 - 2 n_f)$, with $n_f$ the number of active quark flavours, and $\Lambda_{\rm QCD}$ is the RGI scale that characterises the effective charge.

The analogue for the tensor charge is ($C_F = (N_c^2-1)/2/N_c$): 
\begin{equation}
\hat \delta_T^{\mathpzc q} = \lim_{\Lambda_{\rm UV} \to\infty} 
\ln [ \Lambda_{\rm UV}^2/\Lambda_{\rm QCD}^2 ]^{\gamma_m/3} \,  
\delta_T^{\mathpzc q}(\zeta)  {\mathpzc E}_{\mathpzc V}(\Lambda_{\rm UV} ,\zeta ; C_F) ;
\end{equation}
equivalently, 
\begin{equation}
\delta_T^{\mathpzc q}(\zeta)  {\mathpzc E}_{\mathpzc V}(\Lambda_{\rm UV} ,\zeta ; C_F) 
= \delta_T^{\mathpzc q}(\zeta^\prime)  {\mathpzc E}_{\mathpzc V}(\Lambda_{\rm UV} ,\zeta^\prime ; C_F)  ,
\end{equation}
where 
\begin{equation}
\delta_T^{\mathpzc q}(\zeta) = \int_0^1 dx\, [ h_1^{\mathpzc q}(x;\zeta) - h_1^{ \bar {\mathpzc q}}(x;\zeta) ].
\label{deltaTfromDF}
\end{equation}

Regarding Eq.\,\eqref{deltaTfromDF}, it is worth noting that angular momentum conservation precludes a gluon contribution to nucleon transversity.  
Since gluon splitting is the source of sea quark DFs, one might expect no sea (non-valence quark) contribution to nucleon transversity.  
This is the case under AO evolution.

 In the last decade or so, numerous theory studies have sought to determine the proton tensor charges.  
 The most recent continuum and lattice results are presented in Refs.\,\cite{Wang:2018kto, Park:2025rxi}, respectively; see Table~\ref{tensortable}.
 They are consistent within mutual uncertainties; so can be averaged to yield:
 \begin{equation}
 \label{tensoraverage}
 \delta_T^{\mathpzc u}(\zeta_2) = 0.83(2) , 
 \quad  \delta_T^{\mathpzc d}(\zeta_2) = -0.20(1) .
 \end{equation}
We use these as the reference values.  They correspond to the following RGI charges:
\begin{equation}
\label{RGItensor}
\hat \delta_T^{\mathpzc u} = 1.74(5) , 
\quad   \hat \delta_T^{\mathpzc d} = -0.43(2) .
\end{equation}

\begin{table}[t]
\caption{\label{tensortable}
Flavour separated proton tensor charges used herein, listed at $\zeta_2=2$\,GeV.
Regarding Ref.\,\cite{Cocuzza:2023oam}, ``DO'' indicates the fits obtained using only available data and ``LA'' identifies those arrived at after additionally imposing an integral constraint determined by lQCD results for the tensor charges \cite{Gupta:2018lvp, Alexandrou:2019brg}.
(\emph{N.B}.\  To assist in formatting, we list $-\delta_T^d$.)
 }\vspace*{-3ex}
\begin{center}
\begin{tabular*}
{\hsize}
{
l@{\extracolsep{0ptplus1fil}}
|c@{\extracolsep{0ptplus1fil}}
c@{\extracolsep{0ptplus1fil}}
c@{\extracolsep{0ptplus1fil}}
|c@{\extracolsep{0ptplus1fil}}
c@{\extracolsep{0ptplus1fil}}
c@{\extracolsep{0ptplus1fil}}}\hline\hline
 & \cite{Wang:2018kto} & \cite{Park:2025rxi} & Eq.\,\eqref{tensoraverage} $\ $ & \cite[RB]{Radici:2018iag} & \cite[JDO]{Cocuzza:2023oam} & \cite[JLA]{Cocuzza:2023oam}\\\hline
$\phantom{-}\delta_T^u\ $ & $0.87(4)\ $ & $0.78(3)\ $ & $0.83(2)\ $ & $0.39(10)\ $ & $0.50(07)\ $ & $0.71(2)\ $\\
$-\delta_T^d\ $ & $0.21(1)\ $ & $0.20(2)\ $ & $0.20(1)\ $ & $0.11(26)\ $ & $0.02(13)\ $ & $0.20(1)\ $ \\
\hline \hline
\end{tabular*}
\smallskip
\end{center}
\end{table}

\begin{table}[t]
\caption{\label{InterpTransv}
{\sf Panel A}.
Interpolation coefficients for transversity DFs depicted in Fig.\,\ref{FigTransud}A, to be used in Eq.\,\eqref{coeffsTransv}.
Recall, the DF is 
$h_1^{\mathpzc q}(x) = {\mathpzc s}^{\mathpzc q} \tilde {\mathpzc r}^{\mathpzc q} (x) {\mathpzc q}(x)$,
${\mathpzc s}^{\mathpzc u}  = 1 = - {\mathpzc s}^{\mathpzc d}$.
{\sf Panel B}. Interpolation coefficients for CI transversity DFs depicted in Fig.\,\ref{FigTransud}B.
 }\vspace*{-3ex}
\begin{center}
\begin{tabular*}
{\hsize}
{
l@{\extracolsep{0ptplus1fil}}
|c@{\extracolsep{0ptplus1fil}}
c@{\extracolsep{0ptplus1fil}}
c@{\extracolsep{0ptplus1fil}}}\hline\hline
{\sf A} ${\mathpzc u}\ $ & $a_0\ $ & $a_1\ $ & $b_1\ $ \\\hline 
\hspace*{0.7em} upper & $2.22\ $ & $\;24.41\ $ & $\;29.05\phantom{1}\ $ \\
\hspace*{0.7em} central & $1.23\ $ & $\phantom{-}0.95\ $ & $\phantom{-}1.86\phantom{1}\ $ \\
\hspace*{0.7em} lower & $1.11\ $ & $-0.52\ $ & $-0.011\ $ \\
\hline
\hspace*{0.7em} ${\mathpzc d}\ $ & & & \\
\hspace*{0.7em} upper &  $0.66\ $ & $-0.29\ $ & $\phantom{1}0.69\ $ \\
\hspace*{0.7em} central$\ $ & $0.72\ $ & $\phantom{-}0.44\ $ & $\phantom{1}2.48\ $  \\
\hspace*{0.7em} lower & $1.11\ $& $\phantom{-}3.71\ $ & $11.36\ $  \\
\hline
\end{tabular*}
\smallskip

\begin{tabular*}
{\hsize}
{
l@{\extracolsep{0ptplus1fil}}
|c@{\extracolsep{0ptplus1fil}}
c@{\extracolsep{0ptplus1fil}}
c@{\extracolsep{0ptplus1fil}}}\hline
{\sf B} ${\mathpzc u}\ $ & $a_0\ $ & $a_1\ $ & $b_1\ $ \\\hline 
\hspace*{0.7em} upper & $2.30\ $ & $\;25.49\ $ & $\;29.93\phantom{1}\ $ \\
\hspace*{0.7em} central$\ $  & $1.27\ $ & $\phantom{-}1.16\ $ & $\phantom{-}2.06\phantom{1}\ $ \\
\hspace*{0.7em} lower & $1.14\ $ & $-0.58\ $ & $-0.083\ $ \\
\hline
\hspace*{0.7em} ${\mathpzc d}\ $ & & & \\
\hspace*{0.7em} upper &  $0.69\ $ & $-0.30\ $ & $\phantom{1}0.65\ $ \\
\hspace*{0.7em} central$\ $ & $0.74\ $ & $\phantom{-}0.42\ $ & $\phantom{1}2.32\ $  \\
\hspace*{0.7em} lower & $1.07\ $& $\phantom{-}3.20\ $ & $\phantom{1}9.79\ $  \\
\hline\hline
\end{tabular*}
\end{center}
\end{table}

\begin{figure}[t]
\begin{center}
\begin{tabular}{l}
\includegraphics[clip, width=0.42\textwidth]{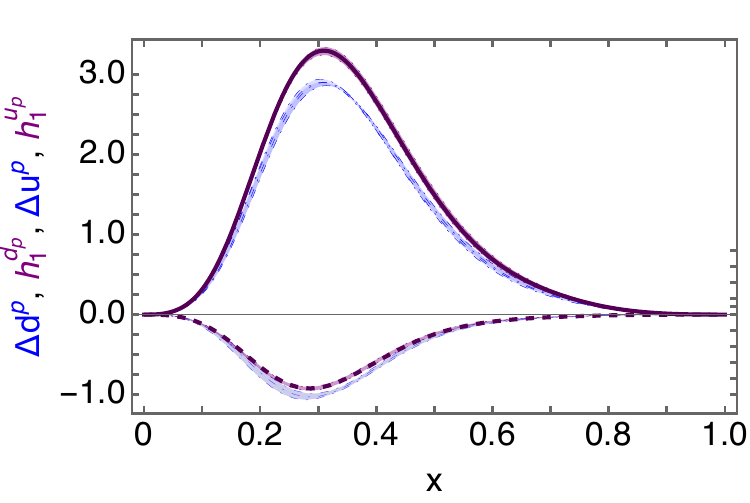} \\[-30ex]
\hspace*{0.39\textwidth}{\large\sf A}\\[26ex]


\includegraphics[clip, width=0.42\textwidth]{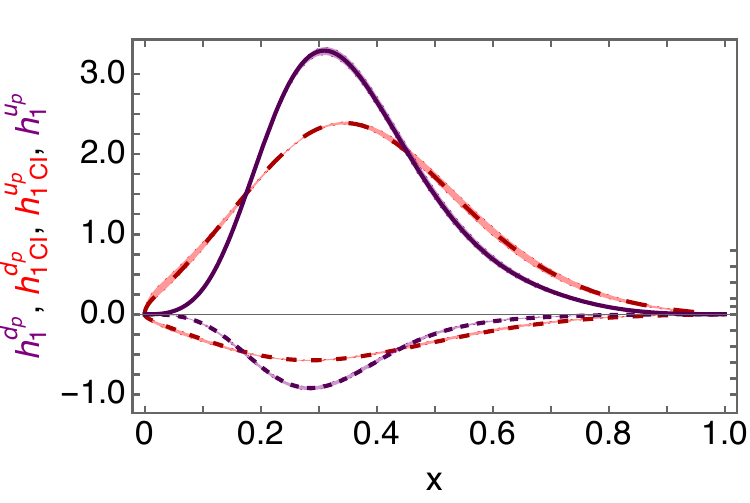} \\[-30ex]
\hspace*{0.39\textwidth}{\large\sf B} 
\end{tabular}
\end{center}
\vspace*{23ex}

%
\caption{\label{FigTransud}
{\sf Panel A}.
Flavour-separated, hadron-scale, in-proton transversity DFs:
$h_1^{\mathpzc u}(x;\zeta_{\cal H})$ -- solid purple curve;
$h_1^{\mathpzc d}(x;\zeta_{\cal H})$ -- dashed purple curve.
The bracketing bands express the uncertainties in Eq.\,\eqref{tensoraverage}.
For comparison, the analogous helicity DFs are drawn as dot-dashed blue curves.
{\sf Panel B}.
Comparison between transversity DFs in Panel~A  and those obtained in the same way, but based on unpolarised proton DFs calculated using a symmetry-preserving treatment of a contact interaction \cite{Yu:2024qsd}:
$h_{1 {\rm CI}}^{\mathpzc u}(x;\zeta_{\cal H})$ -- long-dashed red curve;
$h_{1 {\rm CI}}^{\mathpzc d}(x;\zeta_{\cal H})$ -- dashed red curve.
}
\end{figure}

At this point, the mappings in Eq.\,\eqref{ratiofunctions} can be used to develop pointwise representations of the hadron-scale transversity distributions.  
We refine the Ref.\,\cite{Cheng:2023kmt} procedure somewhat: after obtaining the profile functions, we subsequently identify and smooth the upper and lower boundaries defined by the twelve curves, 
and obtain final results that can be written in the form 
$h_1^{\mathpzc q}(x) = {\mathpzc s}^{\mathpzc q} \tilde {\mathpzc r}^{\mathpzc q} (x) {\mathpzc q}(x)$,
\begin{equation}
\label{coeffsTransv}
\tilde{\mathpzc r}^{\mathpzc q} = \sqrt{x} \, \frac{a_0 + a_1 x}{1+b_1 x} ,
\end{equation}
with the interpolation coefficients listed in Table~\ref{InterpTransv}.  

The hadron-scale transversity DFs are drawn in Fig.\,\ref{FigTransud}A and compared therein with analogous flavour-separated helicity profiles.  
As found in earlier studies \cite{Cloet:2007em, Mondal:2019jdg}, for each flavour, $u$ and $d$, helicity and transversity profiles are similar.
The quantitative differences between peak magnitude maxima and locations owe to the sizes of quark axial charges relative to the hadron-scale tensor charges.

{\allowdisplaybreaks
Splitting functions for helicity and transversity DFs are different:
\begin{align}
& P^{\Delta {\mathpzc q}}(x) = C_F\bigg[ \frac{1+x^2}{(1-x)_+} + \tfrac{3}{2}\delta(1-x) \bigg] \nonumber \\
& \to 
P^{h_1}(x) = C_F\bigg[ \frac{2 x}{(1-x)_+} + \tfrac{3}{2}\delta(1-x) \bigg] . \label{Splitting}
\end{align}
Consequently, the similarities between these DFs displayed in Fig.\,\ref{FigTransud} are lost under evolution. 
The Soffer bound \cite{Soffer:1994ww} is satisfied pointwise at $\zeta_{\cal H}$ and this remains true under evolution. 
}

Figure~\ref{FigTransud}B compares the transversity DFs calculated as described above with those obtained in the same way but beginning with unpolarised proton DFs computed using a symmetry-preserving treatment of a contact interaction (CI) \cite{Yu:2024qsd}.  
Naturally, a CI produces harder DFs than a framework built upon momentum-dependent gluon exchange.
The impact of this difference is apparent in the image: the CI results have greater support on the endpoint domains and give-up support at mid-$x$.  Kindred curves in Fig.\,\ref{FigTransud}B deliver the same hadron-scale tensor charges, listed in Eq.\,\eqref{tensoraverage}.

\begin{figure}[t]
\begin{center}
\begin{tabular}{l}
\includegraphics[clip, width=0.42\textwidth]{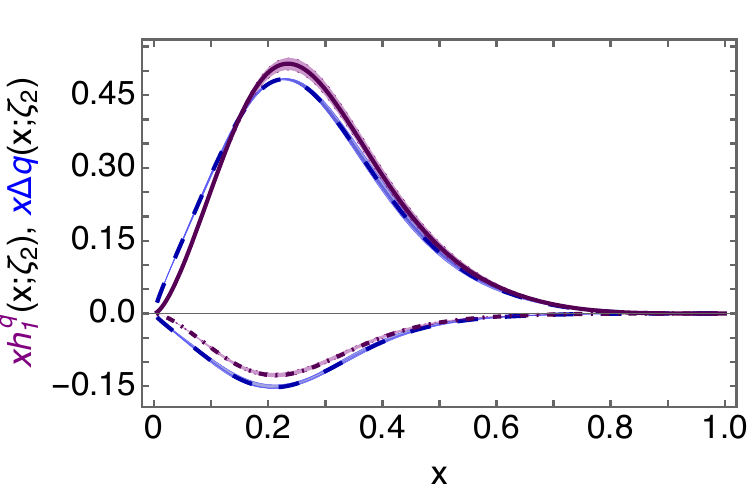} \\[-30ex]
\hspace*{0.39\textwidth}{\large\sf A}\\[26ex]


\includegraphics[clip, width=0.42\textwidth]{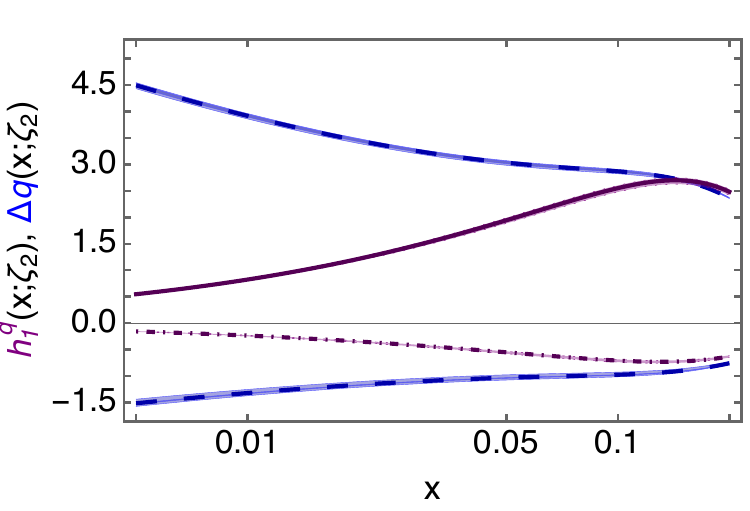} \\[-30ex]
\hspace*{0.39\textwidth}{\large\sf B} 
\end{tabular}
\end{center}
\vspace*{23ex}

%
\caption{\label{FigdcfD2}
{\sf Panel A}.
Flavour-separated, $x$-weighted in-proton transversity DFs:
$xh_1^{\mathpzc u}(x;\zeta_{2})$ -- solid purple curve;
$xh_1^{\mathpzc d}(x;\zeta_{2})$ -- dashed purple curve.
The bracketing bands express the uncertainties in Eq.\,\eqref{tensoraverage}.
For comparison, the analogous helicity DFs are drawn as long-dashed blue curves and bands.
{\sf Panel B}.
Focusing on low-$x$, same transversity and helicity DFs without the $x$-weighting.
}
\end{figure}

\begin{table}[t]
\caption{\label{zeta2fit} 
Polynomial coefficients for use in the representation approximation for transversity DFs described in connection with Eq.\,\eqref{PlotApprox}.
{\sf Panel A} -- based on realistic proton DFs, $\alpha = 0.494$, $\beta = 4.63$; 
{\sf Panel B} -- based on contact interaction proton DFs, $\alpha = 0.325$, $\beta = 3.35$.
 }
 \vspace*{-3ex}
\begin{center}
\begin{tabular*}
{\hsize}
{
l@{\extracolsep{0ptplus1fil}}
|c@{\extracolsep{0ptplus1fil}}
c@{\extracolsep{0ptplus1fil}}
c@{\extracolsep{0ptplus1fil}}}\hline\hline
{\sf A} &  $a_1\ $ & $a_2\ $ & $a_3\ $ \\\hline 
${\mathpzc u}\ $ upper & $12.7\ $& $-39.5\ $ & $31.7\ \ $ \\
$\phantom{\mathpzc u}\ $ central$\ $   & $13.4\ $ & $-42.4\ $ & $34.2\ $ \\
$\phantom{\mathpzc u}\ $ lower &  $ 13.9\ $ &  $-43.8\ $ & $35.1\ $ \\
\hline
 ${\mathpzc d}\ $ upper & $13.4\ $  & $-52.4\ $& $47.4\ $ \\
$\phantom{\mathpzc d}\ $ central$\ $  & $13.1\ $ & $-50.7\ $ & $45.6\ $ \\
$\phantom{\mathpzc d}\ $ lower & $12.0\ $& $-46.8\ $ & $42.2\ $\\
\hline
\end{tabular*}

\smallskip

\begin{tabular*}
{\hsize}
{
l@{\extracolsep{0ptplus1fil}}
|c@{\extracolsep{0ptplus1fil}}
c@{\extracolsep{0ptplus1fil}}
c@{\extracolsep{0ptplus1fil}}}\hline
{\sf B} &  $a_1\ $ & $a_2\ $ & $a_3\ $ \\\hline 
${\mathpzc u}\ $ upper & $2.13\phantom{0}\ $& $-7.39\ $ & $5.18\ \ $ \\
$\phantom{\mathpzc u}\ $ central$\ $   & $2.69\phantom{0}\ $ & $-9.39\ $ & $6.85\ $ \\
$\phantom{\mathpzc u}\ $ lower &  $ 3.10\phantom{0}\ $ &  $-10.7\phantom{00}\ $ & $7.69\ $ \\
\hline
 ${\mathpzc d}\ $ upper & $-0.0840\ $  & $-3.88\ $& $3.73\ $ \\
$\phantom{\mathpzc d}\ $ central$\ $  & $-0.187\phantom{0}\ $ & $-3.17\ $ & $3.10\ $ \\
$\phantom{\mathpzc d}\ $ lower & $-0.780\phantom{0}\ $& $-1.24\ $ & $1.52\ $\\
\hline
\end{tabular*}
\end{center}
\end{table}

\section{Transversity Distributions: 2 GeV}
\label{SecTAO}
After AO evolution of the results in Fig.\,\ref{FigTransud}A ($\zeta_{\cal H}\to \zeta_2$), one obtains the solid and dot-dashed purple curves drawn in Fig.\,\ref{FigdcfD2}.
{\allowdisplaybreaks
For the purpose of ready use elsewhere, these curves are adequately represented using the following functional form:
\begin{subequations}
\label{PlotApprox}
\begin{align}
{\mathpzc f}(x) & = {\mathpzc n}_{\mathpzc f} x^\alpha (1-x)^\beta (1 + a_1 x + a_2 x^2 + a_3 x^3) \\
h_1^{\mathpzc q}(x) & = \delta_T^{\mathpzc q} {\mathpzc f}(x) ,
\end{align}
\end{subequations}
where ${\mathpzc n}_{\mathpzc f}$ ensures $\int_0^1dx {\mathpzc f}(x) = 1$, 
the tensor charges are given in Table~\ref{tensortable}, 
and the exponents and polynomial coefficients are listed in Table~\ref{zeta2fit}.  
}

\begin{figure}[t]
\centerline{\includegraphics[width=0.44\textwidth]{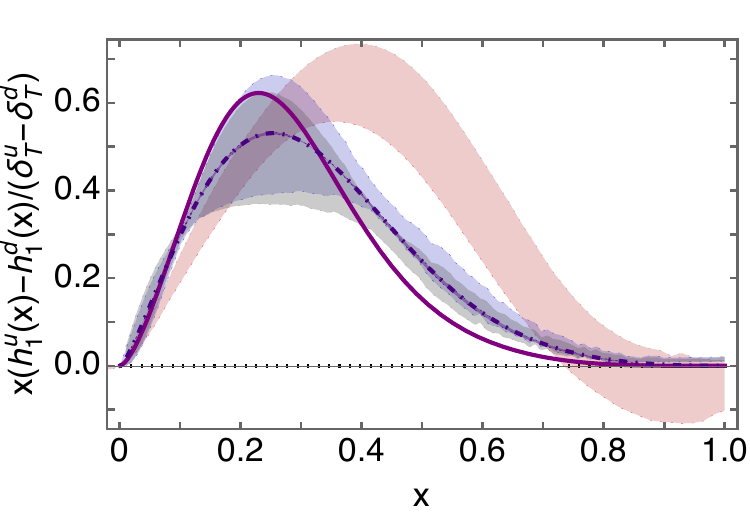}}
\caption{\label{cflQCD}
In-proton isovector tensor charge DF, normalised by its zeroth moment, \emph{viz}.\ the integrated isovector tensor charge:
$h_1^T(x)/g_T$.  (At scale $\zeta_2^2=4\,$GeV$^2$, unless otherwise noted.)
Legend.
Realistic proton DFs -- solid purple curve; 
contact interaction proton DFs -- dot-dashed purple curve.
Ref.\,\cite[HS2]{HadStruc:2021qdf} -- blue band. (This result was computed at $\zeta^2=2\,$GeV$^2$.)
AO evolution of HS2 to $\zeta^2=4\,$GeV$^2$ -- grey band.  (For this evolution, AO is practically equivalent to leading-order evolution.  Four active flavours.) 
Ref.\,\cite[LPC]{LatticeParton:2022xsd} -- red band.
}
\end{figure}

\begin{figure}[t]
\begin{center}
\begin{tabular}{l}
\includegraphics[clip, width=0.42\textwidth]{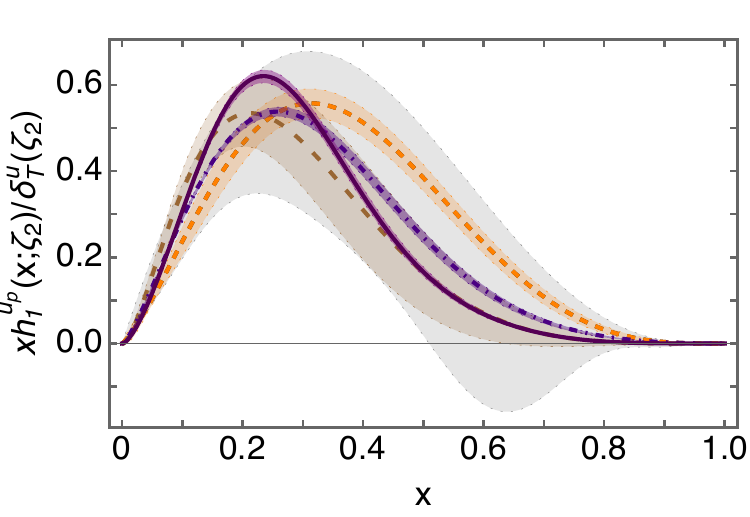} \\[-30ex]
\hspace*{0.39\textwidth}{\large\sf A}\\[26ex]

\includegraphics[clip, width=0.42\textwidth]{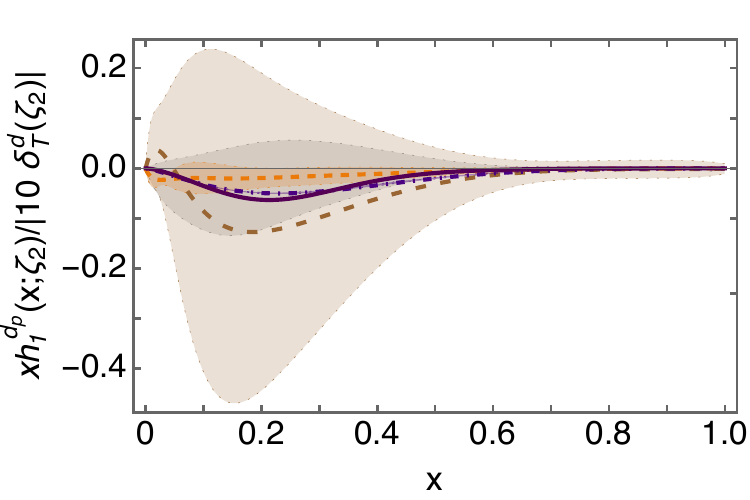} \\[-30ex]
\hspace*{0.39\textwidth}{\large\sf B} 

\vspace*{23ex}

\end{tabular}
\end{center}

%
\caption{\label{FigTransudz2}
Flavour-separated, in-proton transversity DFs at resolving scale $\zeta_2=2\,$GeV.
{\sf Panel A}.
$h_1^{\mathpzc u}(x;\zeta_{2})$ -- solid purple curve; 
$h_1^{{\mathpzc u} {\rm CI}}(x;\zeta_{2})$ -- dot-dashed purple curve.
{\sf Panel B}.
$h_1^{\mathpzc d}(x;\zeta_{2})$ -- solid purple curve;
$h_1^{{\mathpzc d} {\rm CI}}(x;\zeta_{2})$ -- dot-dashed purple curve.
In both cases, the bracketing bands express the uncertainties in Eq.\,\eqref{tensoraverage}.
For comparison, results obtained from phenomenological global fits are also drawn:
\cite[RB]{Radici:2018iag} -- grey band;
\cite[JDO]{Cocuzza:2023oam} -- brown wide-spaced dashed curve and band;
\cite[JLA]{Cocuzza:2023oam} -- orange dashed curve and band. 
\emph{N.B}. To ensure sensible comparisons between all results and as indicated in the ordinate labels, each $u$-quark curve is normalised by its associated tensor charge value and the $d$-quark curves by $10$-times the magnitude of the relevant value. 
}
\end{figure}

Figure~\ref{FigdcfD2} also depicts the helicity DFs in Fig.\,\ref{FigTransud}A as they appear after AO evolution.  
Evidently, if one is solely concerned with $x$-weighted DFs, then even after such evolution, the transversity and helicity profiles are very similar; see Fig.\,\ref{FigdcfD2}A.  
However, owing to Eq.\,\eqref{Splitting}, evolution actually leads to marked differences between the helicity and transversity distributions themselves on $x\lesssim 0.1$; see Fig.\,\ref{FigdcfD2}B: 
helicity DFs must diverge as $x\to 0$; 
but, herein and consistent with expectations \cite{Kovchegov:2018zeq}, transversity DFs vanish as a small power of $x$ in this limit.

A comparison with contemporary lQCD results for the pointwise behaviour of the isovector transversity DF \cite{HadStruc:2021qdf, LatticeParton:2022xsd} is provided in Fig.\,\ref{cflQCD}.  
(lQCD works with this difference in order to avoid difficulties associated with computation of disconnected diagrams.)
Evidently, the lattice results disagree with each other, possibly because they use different ensembles and approaches to DF calculation.  Future studies may reconcile them.  
Notwithstanding, our prediction is consistent with that reported in Ref.\,\cite{HadStruc:2021qdf}. 
Curiously, this lQCD result appears to favour the harder CI prediction over that built upon realistic proton valence quark DFs.
This may change following further studies.

It is worth recalling that the behaviour of DFs on the valence domain, $x\gtrsim 0.25$, is much discussed because inferences from data are very sensitive to the form of perturbative hard scattering kernel used in solving the inverse problem that connects DFs with data.  
Lattice studies are currently only able to deliver reliable results at mid-$x$, \emph{viz}.\ $0.2 \lesssim x \lesssim 0.8$ \cite{Ji:2024hit}.

In Fig.\,\ref{FigTransudz2}, our transversity DF profiles are compared with phenomenological fits \cite{Radici:2018iag, Cocuzza:2023oam}.  
Considering those fits, it is apparent that available data do not provide tight constraints on proton transversity DFs.  
The inclusion of even limited additional information, \emph{e.g}., lQCD estimates of tensor charges, markedly alters the profiles.
Our phenomenologically constructed predictions may therefore be of material value.
This potential is further highlighted by the comparison with contemporary lQCD results for the $x$-dependence of the isovector tensor DF; see Fig.\,\ref{cflQCD}.

\section{Perspective}
We proposed simple \emph{Ans\"atze} for hadron-scale in-proton valence-quark transversity distribution functions (DFs).  
They  were developed using insights from perturbative QCD, analogous helicity DFs, and existing calculations of proton tensor charges, and constrained by solutions of a quark + fully-interacting diquark picture of nucleon structure [Sec.\,\ref{SecT}].
Subsequently exploiting the existence of a QCD effective charge that defines an evolution scheme for parton DFs, we delivered parameter-free predictions for proton transversity DFs at $\zeta_2 = 2\,$GeV [Sec.\,\ref{SecTAO}].  
In doing so, we completed a single-framework unification of all pion and proton DFs that survive in the collinear limit. 

We found that so long as one is only dealing with processes that are sensitive to $x$-weighted transversity DFs at modest resolving scales, it should be possible to obtain reliable estimates by assuming equivalence between the transversity and helicity DF profiles with, perhaps,  rescaling to account for differences between the values of flavour-separated axial and tensor charges [Fig.\,\ref{FigdcfD2}].  
On the other hand, concerning the DFs themselves, the profiles diverge markedly on $x \simeq 0$ owing to differences between the equations governing their scale evolution. 

Regarding results from lattice-regularised QCD (lQCD) [Fig.\,\ref{cflQCD}], there is some tension between the two available profiles for the isovector transversity DF.
Our predictions align better with that in Ref.\,\cite{HadStruc:2021qdf}.  
With the growing interest in DF extraction using lQCD, one may anticipate additional information in the foreseeable future. 

The paucity of data that can today be used for inference of proton transversity DFs means that phenomenological fits are poorly constrained and possess large uncertainties.  
The significance of similarities and differences between our parameter-free results and extant fits is therefore difficult to judge [Fig.\,\ref{FigTransudz2}].
Given the absence of, \emph{e.g}., lQCD results for the pointwise behaviour of the flavour-separated charges, it may be worth using our results to constrain such fits.  
 
The results delivered herein can be tested and improved in future by working with a nucleon tensor current that is appropriate 
for a proton represented as a bound state of dressed-quark and fully interacting nonpointlike diquark degrees-of-freedom, and using that to calculate proton hadron scale transversity DFs.  
First steps along this path are being taken by using the contact interaction framework, with the required subsystem tensor form factors already available \cite{Guo:2026ccd}.  A natural subsequent extension is to advance the analysis using Schwinger functions that possess a closer connection with QCD. 
A longer term goal would see extension of the Poincar\'e-covariant three-body Faddeev equation analysis that delivered proton tensor charges \cite{Wang:2018kto} to calculation of the associated DFs.  

\medskip

\noindent\textbf{Acknowledgments}.
We thank ECT* for hosting the Workshop ``Theoretical Motivations and Empirical Evidence for Diquark Correlations in Hadrons'', during which this work was initiated.
Work supported by:
National Natural Science Foundation of China, grant no.\ 12135007;
Natural Science Foundation of Anhui Province, grant no.\ 2408085QA028; 
Natural Science Foundation of Jiangxi Province, grant no.\ 20262BAC200542;
Early-Career Young Scientists and Technologists Project of Jiangxi Province, grant no.\ 20262BEJ730037.

%
%

\medskip
\noindent\textbf{Data Availability Statement}. This manuscript has no associated data or the data will not be deposited. [Authors' comment: All information necessary to reproduce the results described herein is contained in the material presented above.]

\medskip
\noindent\textbf{Declaration of Competing Interest}.
The authors declare that they have no known competing financial interests or personal relationships that could have appeared to influence the work reported in this paper.


\end{document}